\documentstyle[referee]{mn}
\input{epsf}

\title{Pulsation modes for increasingly relativistic polytropes}

\author[Nils Andersson and Kostas D. Kokkotas]
{ Nils Andersson$^{1}$ and Kostas D. Kokkotas$^{2}$\\ 
$^{1}$ Department of Physics,
Washington University, St Louis MO 63130, USA \\
$^{2}$ Department of Physics, Aristotle University of Thessaloniki,
Thessaloniki 54006, Greece} 

\date{Accepted 1997  (?).
      Received 1997  (?);
      in original form  1997}

\begin{document}

\maketitle

\begin{abstract}
We present the results of a numerical study of the fluid $f$, $p$
and the gravitational $w$ modes for increasingly relativistic
nonrotating
polytropes. The results for $f$ and $w$-modes are in good agreement
with previous data for uniform density stars, which supports an
understanding of  the nature of the gravitational wave modes 
 based on the uniform density
data. We show that the $p$-modes can become extremely long-lived
for some relativistic stars. This effect is attributed to the
change in the perturbed density distribution as the star becomes
more compact. 
\end{abstract}


\section{Introduction}

In a recent paper \cite{akk} we presented a detailed survey of the 
pulsation modes
of the simplest conceivable stellar model: a nonrotating
star with uniform density. 
The reasons for chosing this, admittedly unrealistic, model were 
twofold: First of all, the analytic solution to the 
TOV equations for  uniform 
density is well known \cite{bfsbook}. 
This means that calculations could readily be 
done for many different values stellar compactness and average density. 
Secondly, more
realistic models for a neutron star are ``almost constant density'' so 
the results for this simple model may not be too different from what 
one would find for realistic 
equations of state. 

The main conclusions of our study were: 
\begin{itemize}
\item  that gravitational-wave ($w$) modes \cite{kokkotas92}
exist for all stellar models, even though the modes become
very short lived for less relativistic stellar models.
 Moreover, the $w$-mode spectra 
are qualitatively similar for axial and polar perturbations (for a description
of the two classes of relativistic 
perturbations -- also referred to as odd and even parity 
perturbations -- see Thorne and Campolattaro \shortcite{thorne} or Kojima \shortcite{kojima}). Since axial 
perturbations do not couple to pulsations in the stellar fluid 
\cite{thorne} we concluded that the $w$-modes are pure ``spacetime'' 
modes that do not depend on the particulars of the fluid for their 
existence. This notion is supported by results obtained for 
various model scenarios \cite{ica,namodel}.

\item that the behaviour of 
the various pulsation modes changes dramatically as the star is made 
more compact than $R\approx 3M$ (we use geometrized units $c=G=1$). For 
such ultracompact stars the existence of a peak in the effective 
curvature potential outside the surface of the star will affect the 
various modes. Some modes can basically be considered as ``trapped in a 
potential well'', cf. Chandrasekhar and Ferrari \shortcite{cfaxial}
and Kokkotas \shortcite{kokkotas94}. 
This trapping has the effect that 
some $w$-modes become extremely long-lived. 

\item that there are ``avoided crossings'' between the fluid 
$f$-mode and the various polar $w$-modes for extremely compact stars.
This suggests that the $f$-mode should be considered as the first in the 
sequence of trapped modes for these stars.
\end{itemize}

The results of the mode-survey for uniform density models
helped improve our understanding of the 
origin of, and the relation between, various  pulsation modes 
of relativistic stars. We would expect the main conclusions to hold
also for more realistic equations of state (at least qualitatively),
but the motivation for confirming these expectations by actual
calculations is  strong. 
For example,  axial $w$-modes have so far only 
been calculated for uniform density stars. Although our present 
understanding of the origin of the $w$-modes indicates that 
axial modes must exist for \underline{all} stellar models it is 
important to verify that this is, indeed, the case.  Furthermore, 
more realistic stellar models support other families of modes. Of 
particular interest for gravitational-wave physics are the $p$-modes 
(that correspond to pressure waves in the stellar fluid).
The fact that  the properties
of the $f$- and $w$-modes change dramatically as the star becomes 
increasingly relativistic leads to questions whether the $p$-modes are 
also affected in an interesting way. With this short paper we address
these issues.

We describe results obtained for the polytropic 
equation of state
\begin{displaymath}
p = \kappa \rho^{\Gamma} \ ,
\end{displaymath}
where $\kappa = 100 \mbox{ km}^2$. Our main study was for $\Gamma = 2$,
but we also considered other (especially larger) values of $\Gamma$. 
Our results are basically an extension of those presented by
Andersson, Kokkotas and Schutz  \shortcite{aks}. A detailed description 
of the way that we extract the complex frequencies of the various 
pulsation modes can be found in that paper. In order to keep the present 
paper short, we will discuss neither this numerical approach nor the 
various perturbation equations here. 
The equations governing polar perturbations are described in detail by 
Lindblom and Detweiler \shortcite{ld}, while the axial equations are 
given by Chandrasekhar and Ferrari \shortcite{cfaxial}.  

\section{Numerical results}

\begin{figure}
\centerline{\epsfxsize=10cm \epsfbox{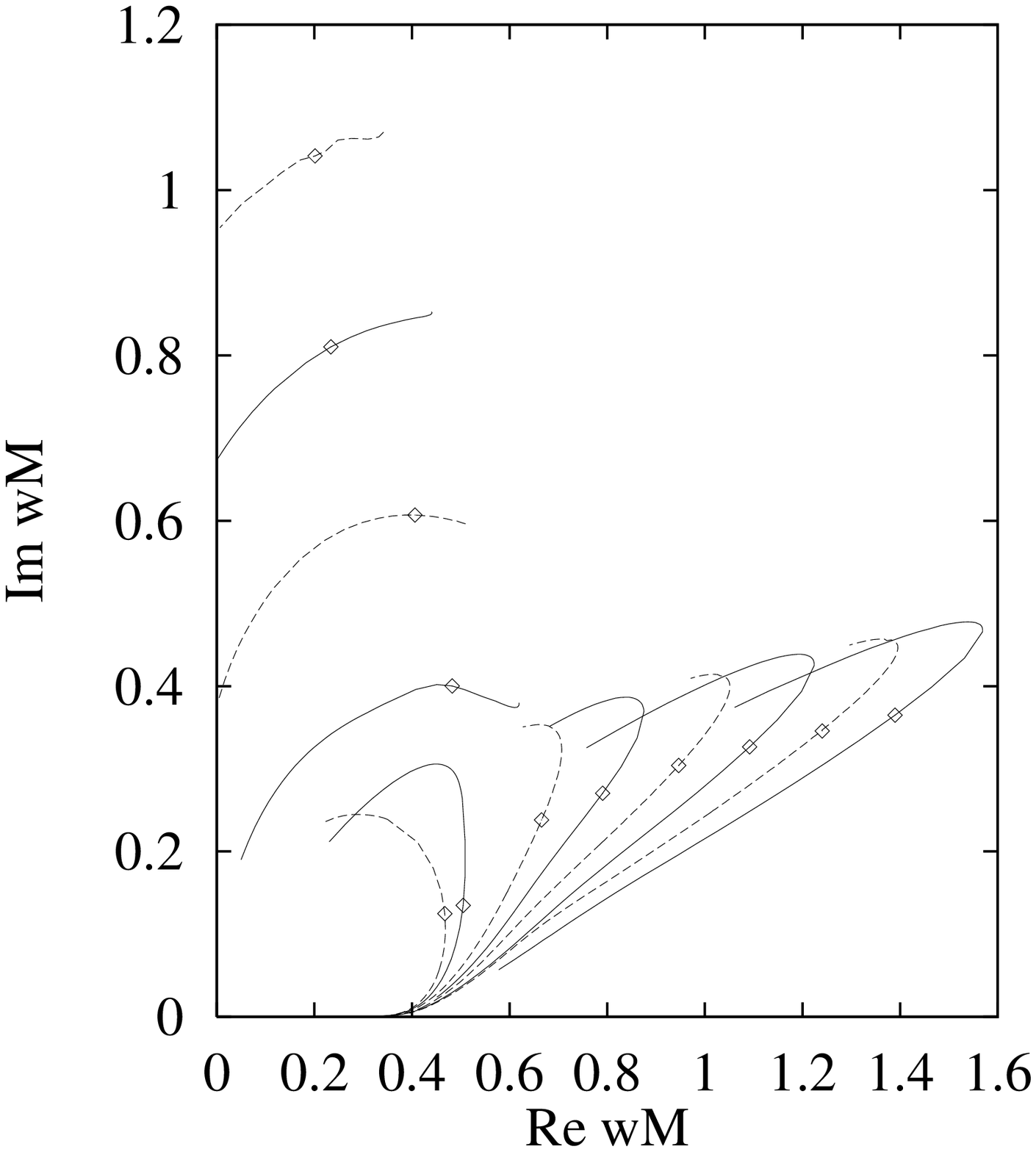}}
\centerline{\epsfxsize=10cm \epsfbox{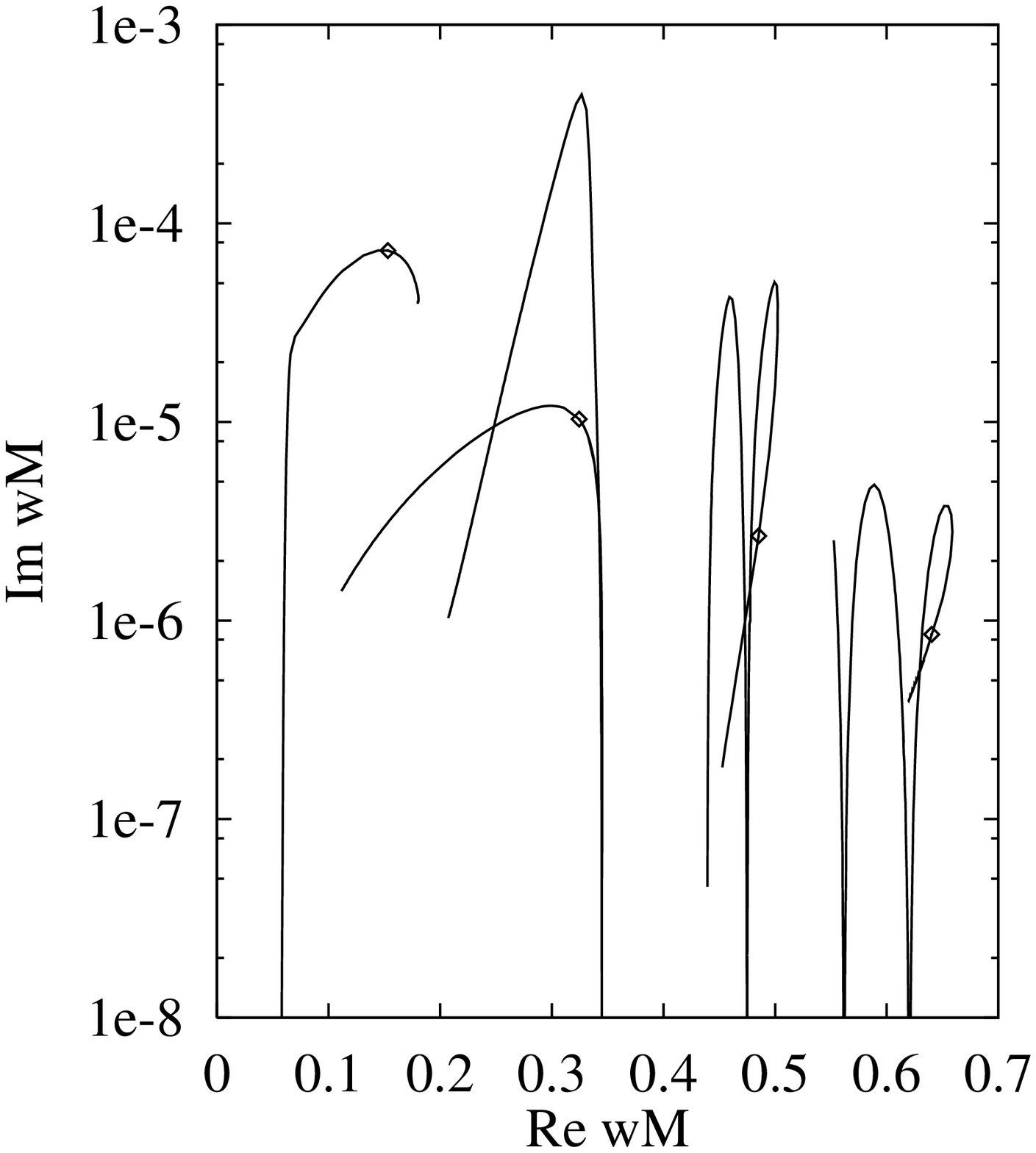}}
\caption{The change in the complex  
mode-frequencies as the central density of the 
stellar model varies ($\rho_c$ increases in the direction of the
arrows). 
The data are for a $\Gamma = 2$ polytrope. The 
gravitational $w$-modes are shown in the upper panel, 
while the much 
longer lived  fluid $f$- and $p$- modes are in the the lower panel
(The
$f$-mode is the leftmost mode in the lower panel.). Axial
$w$-modes are represented by dashed lines while polar modes are shown 
as solid lines. The diamonds on each curve indicates the densest
stellar
model that is stable to radial perturbations.}
\label{fig1}\end{figure}

The results of our investigation for increasingly relativistic
polytropes can 
be summarized in a single figure. In Figure~\ref{fig1} we show the 
inverse damping rate (represented by $\mbox{Im } \omega M$) of each
mode as a function 
of the pulsation frequency ($\mbox{Re }\omega M$)  for a
 $\Gamma =2$ polytrope. As the central density of the 
stellar model is varied each mode traces out a curve in the complex 
$\omega$-plane. In the figure we have indicated (by diamonds) the 
densest polytropic model that is stable to radial perturbations.
That is, the specific model for which the mass ($M$)  reaches a
maximum as a function of the 
central density ($\rho_c$). For the $\Gamma=2$ polytrope used to
calculate
the data in Figure~\ref{fig1}  the marginally stable model
has central density 
$\rho_c = 5.7\times 10^{15} \mbox{ g/cm}^3$. 

It is clear that Figure~\ref{fig1} contains a considerable amount
of information. We now proceed to
discuss the results for each separate class of modes in more detail.

%put figure 1 around here

\subsection{Gravitational wave modes}

As far as the $w$-modes are concerned our investigation does not add
much to the study for uniform density stars \cite{akk}. 
The results for polytropes are, both qualitatively and quantitatively,
in excellent agreement with those for the uniform density model. 
From the upper panel in Figure~\ref{fig1} it is clear that both the axial
and the polar $w$-modes can be divided into two separate families. The 
two families are distinguished by the behaviour of the 
mode-frequencies as the star becomes very relativistic. For one family 
of modes -- the interface $w$-modes \cite{akk,nollert} -- the frequencies 
approach constant (complex) values for large $\rho_c$. For the 
specific polytrope under consideration we find that the mode-frequencies 
hardly change at all
for $\rho_c > 1.9\times 10^{16} \mbox{ g/cm}^3$. The modes in 
the second family -- the curvature $w$-modes -- are characterized by
an increase in $|\omega M|$  with $\rho_c$ for less compact stars, but
before 
the star becomes radially unstable $|\omega M|$ reaches a maximum
value. 
The modes in this family then
become extremely slowly damped as $\rho_c$ increases further.
The slowest damped $w$-modes become long-lived  as $R<3.5M$ or so.
The decreased damping rate 
is most likely due to the increased influence of the curvature 
potential barrier (the 
peak of which is at $R\approx 3M$). However, it is important
to realize that the $\Gamma=2$ 
polytrope that we consider is always less compact than $R=3.2M$. Hence, 
the surface of these stars is never located inside the peak of the 
curvature potential. That the modes still become extremely slowly
damped  shows that the exterior potential can ``trap''
gravitational waves effectively even though there is no
``potential well inside a barrier'' for these stars, cf. the 
discussion by Detweiler \shortcite{detweiler}.

As already mentioned, the present results are very similar to those 
for uniform density stars. Especially worth noticing are the 
similarities between the axial and the polar $w$-modes
in Figure~\ref{fig1}. Since the axial modes cannot
induce oscillations in the stellar fluid, this indicates that the
character of the $w$-mode depends solely on the properties of
the curved spacetime.
In other words: the $w$-modes are ``spacetime'' modes 
the details of which do not depend on the dynamics of the stellar
fluid \cite{akk,ica}. 

\subsection{Fluid modes}

The present results for the fluid $f$-mode are also in good 
agreement with the uniform density study \cite{akk}. 
For less compact stars the 
damping rate of the $f$-mode increases
with $\rho_c$. This simply means that the star radiates  
gravitational waves more efficiently as it becomes more
compact. But, in a similar way to the 
curvature $w$-modes, the $f$-mode becomes slower damped once the 
star approaches  $R=3M$. Again, this can be understood as the
increasing influence of the curvature potential \cite{detweiler,akk}.

A notable difference between the results for 
the $\Gamma =2$ polytrope and the uniform density star is the 
absence of ``avoided crossings'' between the polar $w$-modes and the 
$f$-mode, cf. Figure 2 of Andersson, Kojima and Kokkotas
\shortcite{akk}.
 The reason for this is that the $\Gamma=2$ 
polytrope never becomes sufficiently compact for such avoided 
crossings to occur.
For the uniform density model the first avoided crossing occured for 
$R \approx 2.3M$, and for the $\Gamma=2$ polytrope we always have 
$R>3.2M$.
One would, however, expect avoided mode-crossings to exist 
for polytropes that can be made very compact. We have verified
the existense of avoided crossings for 
a $\Gamma = 5$ model (which can be made as compact as $R\approx
2.39M$). 
The existence of avoided crossings thus seems to be a generic feature
of extremely compact relativistic stars.

The results we present for $p$-modes are both new and 
remarkable. 
Fluid $p$-modes have not previously been calculated for extremely 
relativistic stellar models. 
As can be seen in Figure~\ref{fig1} these modes behave in a  
peculiar ways as $\rho_c$ increases. While the behaviour for less 
relativistic stars is analogous to the $f$- and the $w$-modes -- 
the damping rate increases when the star becomes more compact -- 
it is different
for the very relativistic models. Once the star has become 
radially unstable $|\omega M|$ attains a maximum.  Then the pulsation 
frequency $\mbox{Re }\omega M$ typically decreases monotonically. At 
the same time the damping rate decreases drastically and 
$\mbox{Im }\omega M$ has a sharp minimum.
For some modes we find several such minima of $\mbox{Im }\omega M$, 
cf. Figure~\ref{fig1}.
The corresponding values of  the 
imaginary part of the mode-frequency are actually so small that
our code does not have sufficient precision to distinguish them from 
zero. It is, of course, important to establish that the sign of 
$\mbox{Im }\omega M$ does not change. A change in sign would
indicate that the mode becomes linearly  unstable. We believe
that our calculations were sufficently accurate to establish that this does 
not happen -- the $p$-modes are all stable. But it is interesting to note
that they become extremey long-lived (very poor radiators of
gravitational waves) for some values of the
central density.

\begin{figure}
\centerline{\epsfxsize=10cm \epsfbox{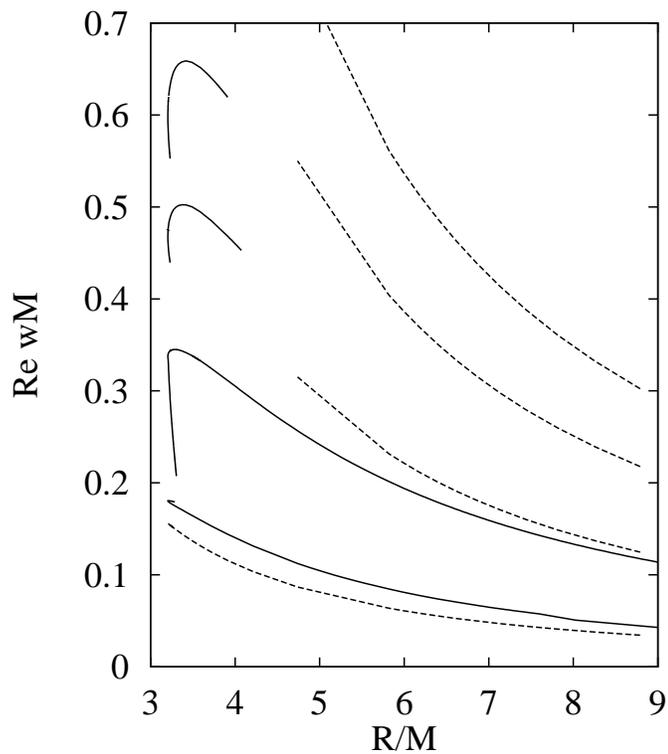}}
\caption{The pulsation frequencies of the fluid modes as functions of
the stellar compactness $R/M$. We compare our results for $f$, $p_0$,
$p_1$ and $p_2$ (solid lines, from bottom to top) to approximate results
for homogeneous stellar models (dashed lines).  }
\label{fig2}\end{figure}

How can we understand the peculiar behaviour of the $p$-mode
frequencies?  Let us first consider the pulsation frequency
$\mbox{Re }\omega M$, and compare our results to ones for simpler
stellar models. The well-known result for the $f$-mode,
established for incompressible fluid spheres 
by Kelvin in 1863 \cite{tassoul} , suggests that our mode should
approach
\begin{equation}
\omega_f^2 = {2l(l-1) \over 2l+1} \left( {M\over R^3} \right) \ ,
\end{equation}
for less relativistic stars.
A similar result, for compressible homogeneous spheres \cite{tassoul},
shows that the $p$-modes ought to approach
\begin{equation}
\omega_p^2 = [ \Delta_n + \sqrt{ \Delta_n^2 + l(l+1) }  ] \left( {M\over
R^3} \right) \ ,
\end{equation}
where
\begin{equation}
2\Delta_n = [2l+3+n(2n+2l+5)]\Gamma -4 \ , \quad n=0,1,2,...
\end{equation}
and $\Gamma$ is the polytropic index. 
As can be seen from Figure~\ref{fig2} our numerical results agree
quite well with these approximations. That is, the pulsation frequency
of each fluid mode changes in the anticipated way as the star
become more relativistic. Alternatively, the results in
Figure~\ref{fig2} can be taken as justification for using the
approximate results also for relativistic stars. They clearly 
provide useful estimates also for objects of neutron star compactness,
$R/M\approx 5$. 
 
%put figure 2 here

The peculiar behaviour of the $p$-mode damping rates can be explained
by changes in the matter distribution inside the star.
According to the standard quadrupole formula, gravitational waves 
due to the fluid motion in the star can be estimated by
\begin{equation}
h \sim  {1\over r} \int_0^R r^4 {\partial^2  \over \partial t^2 } 
\delta\rho (r,t) \ dr =
- {\omega^2 e^{i\omega t}\over r} \int_0^R r^4  \delta\rho (r,\omega)  dr \ ,
\end{equation}
where $\delta \rho$ is the Eulerian variation in the density. 
We have assumed that the
perturbation is monochromatic with a harmonic time-dependence
$\delta \rho(r,t) =  \delta\rho(r,\omega) e^{i\omega t} $.
In the notation of Lindblom and Detweiler \shortcite{ld} the required 
density perturbation follows from (for the quadrupole) 
\begin{equation}
\delta\rho(r,\omega) = - r^2 \left[ e^{-\nu/2} X - {p+\rho \over
r^3} e^{\lambda/2} (M+4\pi pr^3) W - 
  (p+\rho) H_0 \right] { \rho^{1-\Gamma} \over \kappa \Gamma} \ ,
\label{euler}\end{equation}
where we have used the fact that the acoustic wave speed in a polytrope 
is $\kappa \Gamma \rho^{\Gamma-1}$.

To get to a useful result, we must also figure out how each complex-frequency 
mode contributes to the relevant physical quantity 
(such as the density variation). 
The reason is obvious: For complex frequencies 
$\omega$ the $\delta\rho(r,\omega) $ 
that follows from 
(\ref{euler}) will be complex, whereas the corresponding physical quantity
should be real valued. The analysis of this problem proceeds exactly
as for the analogous black-hole problem \cite{testfield}. 
The main steps are i) note that
the modes come in pairs $\omega_n$ and $-\omega_n^\ast$  (where the asterisk 
denotes complex conjugation) ii) Since all perturbation 
equations [see Lindblom and Detweiler \shortcite{ld}] contain only 
$\omega^2$, and $(\omega^2)^\ast = (-\omega^\ast)^2$, we can conclude that
\begin{equation}
 \delta\rho(r,-\omega_n^\ast) = [ \delta\rho (r,\omega_n)]^\ast \  .
\end{equation}
From a specific mode frequency $\omega_n$ we therefore get a contribution (for more details see Andersson \shortcite{testfield})
\begin{equation}
\delta \rho (r,t) \sim 2 \mbox{Re } \left[ \delta\rho (r,\omega_n) 
e^{i\omega_n t} \right] \ .
\end{equation}
Both the real and the imaginary part of the eigenfunction $\delta\rho (r,\omega_n)$ 
will thus be important.

\begin{figure}
\centerline{\epsfxsize=10cm \epsfbox{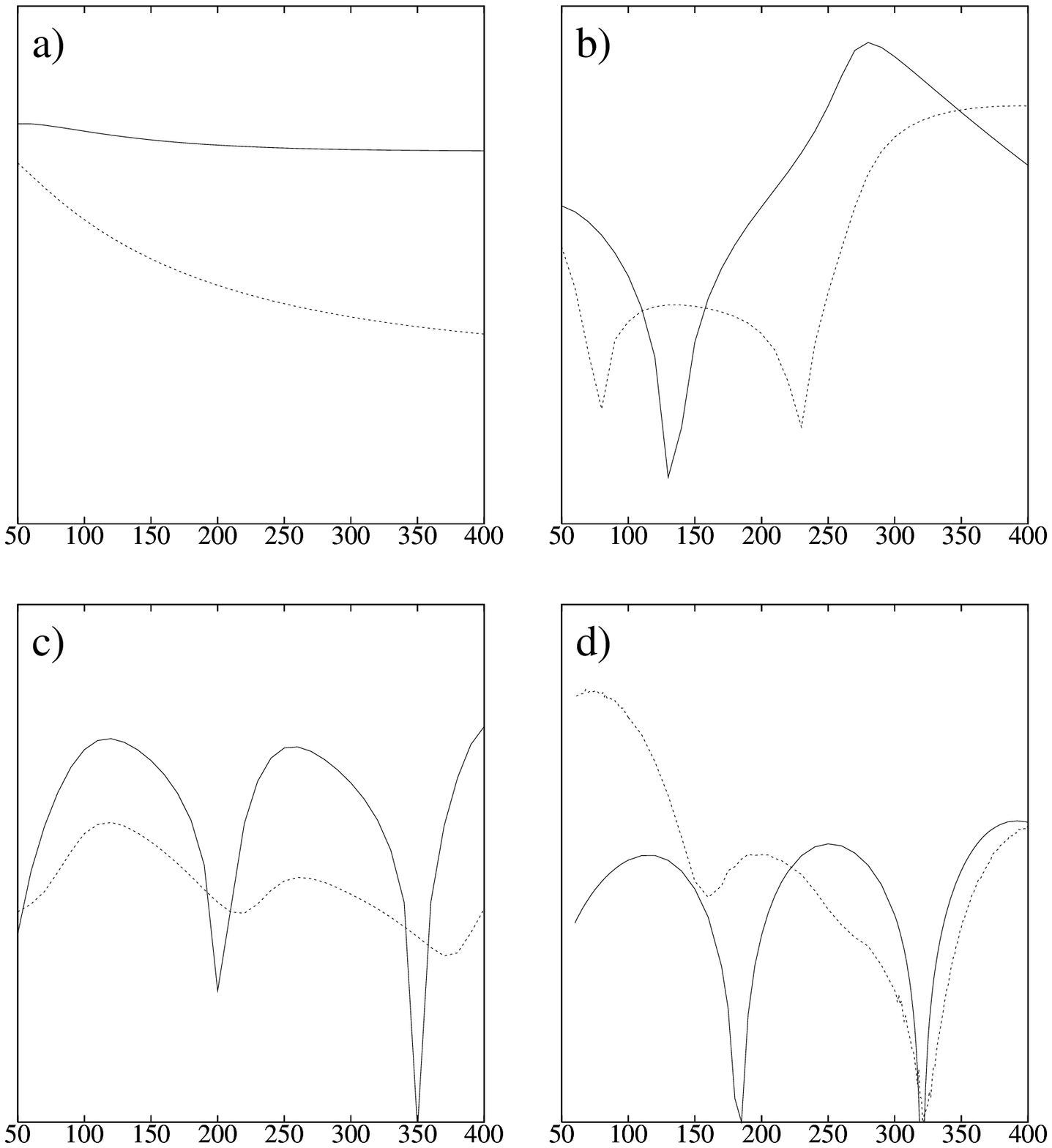}}
\caption{We compare the damping rate of the fluid pulsation modes
(represented by $\mbox{Im }\omega M$ and solid curves) to the efficieny
measure $\epsilon(\omega_n)$ (dashed curves). Both quantities 
are shown as functions
of the central density of the stellar model 
(in units of $10^{14} \mbox{ g/cm}^3$) The scale for $\mbox{Im }\omega
M$ is logarithmic and ranges from $10^{-8}$ to $10^{-2}$ while the scale
for $\epsilon(\omega_n)$  is arbitrary. The data are for a) the $f$-mode,
b-d) the first three $p$ modes. Of interest here is a possible correlation 
between minima in the two functions. Such a correlation explains the
minima in the damping rate of the $p$-modes in terms of changes in the
perturbed
density distribution in the star. (The $f$-mode is only included 
for comparison -- there should be no minima in $\epsilon(\omega_n)$ for
that mode.)
}
\label{fig3}\end{figure}

Let us now assume that the gravitational-wave damping is sufficiently 
slow that the mode is essentially undamped for an entire cycle
(this is a reasonable assumption for the slowly damped fluid modes). 
Averaging over one period we get 
\begin{equation}
<h^2> \sim  e^{-2 {\rm Im } \omega_n t} \left|  \int_0^R r^4  
\delta\rho (r,\omega_n)  dr \right|^2  =  e^{-2 {\rm Im } \omega_n t} \epsilon(\omega_n) \ .
\end{equation}
The value of $\epsilon(\omega_n)$ provides a measure of how ``efficient'' 
a specific mode is as radiator according to the quadrupole formula.
An association between the $p$-mode minima of $\mbox{Im }\omega M$ 
and minima in $\epsilon(\omega_n)$ would indicate that
the features seen in Figure~\ref{fig1} are due to changes in the
perturbed density distribution as $\rho_c$ increases.
In Figure~\ref{fig3} we compare $\mbox{Im }\omega M$ to $\epsilon(\omega_n)$.
This figure seems to establish the anticipated correlation between 
minima in the two functions for the first few $p$-modes.

Having found the likely explanation for the extremely slow damping
of various $p$-modes for some values of $\rho_c$, 
it is worthwhile to discuss how the
eigenfunction $\delta \rho(r,\omega_n)$ changes
as we vary $\rho_c$. Let us first consider the $f$-mode.
This mode is distinguished by the fact that the eigenfunctions 
i) have no nodes inside the star and ii) grow monotonically 
towards the surface of the star. The $p$-modes are different; 
in general there are $n$ nodes in both the real and the imaginary part
of $\delta\rho(r,\omega)$ for the $n$th $p$-mode. 
But the location of these nodes change
 as we vary $\rho_c$. We find that the distribution of $\delta\rho(r,\omega)$
 changes drastically with varying $\rho_c$. Specifically, for the 
 first $p$-mode and $\rho_c < 1\times 10^{16} \mbox{ g/cm}^3$ the single 
 node is close to the surface of the star and the bulk of
$\delta\rho(r,\omega)$ is located in the outer 50\% of the star. But for
$\rho_c > 3\times 10^{16} \mbox{ g/cm}^3$ the node has moved close
to the centre of the star and the bulk of $\delta\rho(r,\omega)$
is now located inside half the radius of the star.  It 
is understandable
that these two, very different, configurations can be rather different
as radiators of gravitational waves. For all $p$-modes the trend is that
the bulk of the eigenfunctions move towards
the centre of the star as we increase $\rho_c$. In a way, this
means that the $p$-modes for the very relativistic star have
the same features as the $g$-modes \cite{tassoul}.

%put figure 3 here

\section{Final comments}

We have presented numerical results for the pulsation modes of
nonrotating relativistic polytropes. These results agree well, 
both qualitatively and quantitatively, with previously obtained
results for uniform density stars \cite{akk}. This provides further
evidence for our understanding of the nature of the various pulsation
modes, especially the notion that the gravitational wave $w$-modes
are mainly due to the properties of the curved spacetime \cite{ica}.
Our present survey also considers the $p$-modes of a relativistic star. 
We find that the polytrope $p$-mode frequencies are close
to an approximation based on the pulsation of a compressible
homogeneous sphere for less relativistic stars. But we
 also show that the $p$-modes change in a peculiar way as the 
star becomes more compact. Specifically, we find that each $p$-mode
can be very slowly damped (in fact, almost undamped) for some 
values of the central density. We have shown that this feature can be 
understood in terms of the change in the distribution of the perturbed
density with varying central density. Although the result is not
of tremendous astrophysical importance (since the affected stars are not
stable to radial perturbations, anyway) it is interesting to note that the
effectivity with which a $p$-mode radiates gravitational waves 
can vary considerably between two stellar models that have almost 
identical central densities.

\pagebreak

\pagebreak

\pagebreak

\begin{thebibliography}{10}

\bibitem[\protect\citename{Andersson} 1996]{namodel}
Andersson N., 1996,  {\em Gen. Rel. Grav} {\bf 28} 1433

\bibitem[\protect\citename{Andersson} 1997]{testfield}
Andersson N., 1997,  {\em Phys. Rev. D} {\bf 55} 468 

\bibitem[\protect\citename{Andersson, Kojima \& Kokkotas }1996]{akk}
Andersson N., Kojima Y., Kokkotas  K.D., 1996,  {\em Ap. J.} {\bf 462} 855

\bibitem[\protect\citename{Andersson, Kokkotas \& Schutz }1995]{aks}
Andersson N.,  Kokkotas  K.D., Schutz B.F, 1995, {\em MNRAS} {\bf 274} 1039

\bibitem[\protect\citename{Andersson, Kokkotas \& Schutz }1996]{ica}
Andersson N.,  Kokkotas  K.D., Schutz B.F, 1996,  {\em MNRAS} {\bf 280} 1230

\bibitem[\protect\citename{Chandrasekhar and Ferrari }1991]{cfaxial}
Chandrasekhar S., Ferrari V., 1991, {\em Proc. R. Soc. Lond. } {\bf
A434} 449

\bibitem[\protect\citename{Detweiler }1975]{detweiler}
Detweiler S.L., 1975   {\em Ap. J.} {\bf 197} 203

\bibitem[\protect\citename{Kokkotas and Schutz }1992]{kokkotas92}
Kokkotas K.D. and Schutz B.F., 1992 {\em MNRAS}
{\bf 255} 119

\bibitem[\protect\citename{Kokkotas }1994]{kokkotas94}
Kokkotas K.D., 1994  {\em MNRAS} {\bf 268} 1015

\bibitem[\protect\citename{Kojima }1992]{kojima}
Kojima Y., 1992, {\em Phys. Rev. D} {\bf 46} 4289 


\bibitem[\protect\citename{Lindblom \& Detweiler }1981]{ld}
Lindblom L., Detweiler S., 1983, {\em Ap. J. Suppl.}  {\bf 53} 73

\bibitem[\protect\citename{Leins, Nollert and Soffel }1993]{nollert}
Leins M., Nollert H-P. and Soffel M.H.,  1993 {\em Phys. Rev. D}
{\bf 48} 3467

\bibitem[\protect\citename{Schutz }1985]{bfsbook}
Schutz B.F., 1985 {\em A first course in general relativity}
 Cambridge Univ. Press 
 
\bibitem[\protect\citename{Tassoul }1978]{tassoul}
Tassoul J.L, 1978 {\em Theory of rotating stars} Princeton Univ. Press

\bibitem[\protect\citename{Thorne \& Campolattaro }1967]{thorne}
Thorne K.S., Campolattaro A., 1967, {\em Ap. J.} {\bf 149} 
591
\end{thebibliography}
\end{document}